# Magneto-Induced Topological Phase Transition in Inverted InAs/GaSb Bilayers

Zhongdong Han,[1,*,#] Tingxin Li,[1,**] Long Zhang,[2] and Rui-Rui Du[1,3,#]

[1]International Center for Quantum Materials, School of Physics, Peking University, Beijing 100871, China

[2]Kavli Institute for Theoretical Sciences and CAS Center for Excellence, University of Chinese Academy of Sciences, Beijing 100190, China

[3]Collaborative Innovation Center of Quantum Matter, Beijing 100871, China

[*] *Present address: Laboratory of Atomic and Solid State Physics, Cornell University, NY 14850, USA.*

[**] *Present address: Key Laboratory of Artificial Structures and Quantum Control (Ministry of Education), School of Physics and Astronomy, Shanghai Jiao Tong University, Shanghai, 200240, China.*

[#] Email: zh352@cornell.edu

Email: rrd@pku.edu.cn

## Abstract

We report a magneto-induced topological phase transition in inverted InAs/GaSb bilayers from a quantum spin Hall insulator to a normal insulator. We utilize a dual-gated Corbino device in which the degree of band inversion, or equivalently the electron and hole densities, can be continuously tuned. We observe a topological phase transition around the magnetic field where a band crossing occurs, that is accompanied by a bulk-gap closure characterized by a bulk conductance peak (BCP). In another set of experiments, we study the transition under a tilted magnetic field (tilt angle $\theta$). We observe the characteristic magneto-conductance around BCP as a function of $\theta$, which dramatically depends on the density of the bilayers. In a relatively deep-inversion (hence a higher density) regime, where the electron-hole hybridization dominates the excitonic interaction, the BCP grows with $\theta$. On the contrary, in a shallowly-inverted (a lower density) regime, where the excitonic interaction dominates the hybridization, the BCP is suppressed indicating a smooth crossover without a gap closure. This suggests the existence of a low-density, correlated insulator with spontaneous symmetry breaking near the critical point. Our highly controllable electron-hole system offers an ideal platform to study interacting topological states as proposed by recent theories.

***Introduction***.—The quantum spin Hall insulator (QSHI) is a topologically non-trivial state described by a $Z_2$ topological invariant [1]. It supports a bulk gap and disspationless helical states along the edge. Several candidate materials have been experimentally confirmed to host the QSH phase, including HgTe/CdTe [2] and InAs/GaSb quantum wells (QWs) [3], 1T'-WTe$_2$ monolayer [4], and potentially transition metal dichalcogenide moiré systems [5]. The physics of the QSH states can be well captured with the Bernevig-Hughes-Zhang (BHZ) model [6] or the Kane-Mele model [7] in the single-particle paradigm. However, an observation in InAs/GaSb bilayer [8]—the persisting conductance quantization up to 12 T in-plane magnetic field—challenges the single-particle picture, which catches more attention on many-body interaction in such systems and opens up a new opportunity to study topological excitonic insulators [9]. Recent theoretical works [10–13] show that topology, correlation, and their interplay can lead to an abundance of exotic states of matter unexplored so far, with the underlying physics being described by correlation-driven spontaneous symmetry breaking. As is common in condensed matter systems, the resulting phase diagram becomes richer and more interesting as an electron-hole bilayer is subject to a magnetic field. Despite some previous experiments under magnetic fields as reported in e.g., [2,14], no attention was paid to the evidence of Coulomb interaction in those reports. Part of the reason may be that in those deeply-inverted InAs/GaSb and HgTe/CgTe QWs studied, the electron-hole (e-h) hybridization dominates and the Coulomb interaction U is nearly negligible compared to the tunneling strength A [15,16]. However, because of the dominance of excitonic interaction over the hybridization, shallowly-inverted InAs/GaSb QWs provide an ideal platform to study those correlation-induced phases. In this Letter, we demonstrate topological phase transitions from a QSHI to a normal insulator (NI) in an inverted InAs/GaSb bilayer under either a perpendicular or a tilted magnetic field.

***Device and measurement***.—InAs/GaSb bilayer is a broken-gap type-II heterostructure where the InAs conduction band is lower than the GaSb valence band by about 150 meV. Intuitively, a straightforward approach to realize a phase transition from a QSHI to a NI is engineering the band crossing. The perpendicular magnetic field is an effective knob for tuning the band alignment due to the orbital effect. Without considering the Coulomb interaction, the band crossing transition should be a topological phase transition accompanied by a bulk gap closure between these two insulating states. On the other hand, in the dilute limit [9] the reduced screening makes the Coulomb interaction dominate over the hybridization. Dual-gated geometry offers *in-situ* and continuous tunability of the effective interaction strength (U/A), which can also be tuned by applying an in-plane magnetic field.

Here we focus on low-temperature bulk transport properties measured from a dual-gated Corbino device, which excludes the edge contribution [8]. Contrary to a Hall mesa, only the bulk contribute to the conductivity which enables us to study the mageto-transport of the system purely from the bulk. Our high-quality bilayer samples are grown by semiconductor molecular beam epitaxy technique. The detailed structure can be found in Supplemental Material. Dual-gated configuration makes the system highly controllable, allowing independent tuning of both the Fermi energy and the band inversion.

Our measurements are performed in $^3$He refrigerators. All data are taken at 300 mK unless specified otherwise. Standard pseudo-four-terminal methods with low-frequency (13.7 Hz) and low-excitation (1 mV) lock-in techniques are adopted in data collection.

***Phase diagram under perpendicular B field.***—Figure 1 shows band structures under different conditions calculated by the two-band BHZ model [6,17]. The electron and hole effective masses are taken from experiments [18]. The crossing density $n_{cross}$ = 1.63 × 10$^{11}$ cm$^{-2}$, which characterizes the band inversion $E_g = \pi\hbar^2 n_{cross}(1/m_e^* + 1/m_h^*)$, is estimated from the analysis of the in-gap oscillation [19]. The bulk inversion asymmetry (BIA) and structure inversion asymmetry (SIA) are not considered here because those effects are normally small [20].

Figure 1(a) and 1(b) present the band dispersion and Landau level (LL) spectra for an inverted case. It represents a simple semimetal, as depicted by dashed lines in Fig. 1(a), if we turn off the band hybridization. Under a uniform magnetic field, there exists two sets of LLs. The perpendicular magnetic field leads to a relative shift of the electron and hole bands in energy, tuning the band structure from an inverted order to a normal (non-inverted) order. This conclusion remains valid even after the hybridization is resumed. The difference is that the hybridization converts the semimetal into a topologically non-trivial QSHI at a zero magnetic field. As shown in Fig. 1(b), it leads to anti-crossings of electron and hole LLs except the lowest Landau levels (LLLs) [16,21]. The only remaining crossing point determined by these two unhybridized LLs reveals a topologically protected gap closure during the phase transition from a (topologically nontrivial) QSHI to a (topologically trivial) NI. Though the time-reversal symmetry is broken in high magnetic fields, this topological phase transition is still protected by other symmetries, such as the spatial mirror symmetry [22,23] due to the vanishingly small BIA and SIA in the InAs/GaSb system [20].

For comparison, for a normal band structure with a non-inverted order—the conduction band is higher in energy than the valence band—the topology is trivial. As shown in Fig. 1(c) and 1(d), the magnetic field pulls the electron and hole LLs apart from the charge neutrality point (CNP) but neither changes the band order nor results in a phase transition.

***Topological phase transition from a QSHI to a NI.*** —Figure 2(a) shows the bulk conductivity as a function of the front-gate voltage $V_f$ and the magnetic field $B_\perp$ at the back-gate voltage $V_b = 0$ V. The conductance shows a dip at $V_f = -0.75$ V under zero magnetic field, indicating the position of the CNP. When fixing the Fermi energy at the CNP, the conductivity in Fig. 2(b) shows clear quantum oscillations, which are followed by a pronounced, broad BCP around 12 T. The in-gap quantum oscillations have been thoroughly discussed in ref. [19] which is a common feature in inverted band structure in InAs/GaSb bilayers and related materials [24–26].

Referring to Fig. 1(b), at the critical B field where the two unhybridized LLs cross, the bulk gap collapses and then a metallic conduction peak emerges, manifesting a topological phase transition from a QSHI to a NI. Two accompanying integer quantum Hall (IQH) states (ii) and (iii) as described in Fig. 1(b) are also observed, which have the opposite Chern numbers as shown by the blue and red dashed lines in Fig. 2(a). They merge into the standard QH sequence at high magnetic field and their Chern numbers (the filling factors) can be determined by invoking the Streda formula [27]. (More detailed information can be found in SM.) A question one may reasonably ask is whether the BCP at 12 T coincides with the in-gap oscillations. A density relevant to the oscillations, $n_{cross} = 1.63 \times 10^{11}$ cm$^{-2}$, can be used to estimate the crossing field of the LLs: $B_c = 2n_{cross}h/e = 13.5$ T. This value is roughly consistent with that of observed 12 T, although several factors may be considered. First, the non-parabolic band dispersions of both bands in a high magnetic field may contribute to the discrepancy. Perhaps more interestingly, deviation from the predicted critical field (13.5 T) may result from the quasi-metallic regime due to e-h weak coupling in InAs/GaSb bilayer and the disorder effect [16].

Furthermore, tracing the conductance peak in Fig 2(a), a "real" peak (rather than, *e.g.*, a saddle point), which peaks not only along the magnetic field axis but also the gate voltage axis, is captured as marked by the red dashed box. It indicates that the bulk gap closes only at one point over the whole gap region in the phase diagram. These observations are fully consistent with a topological phase transition from a QSHI to a NI depicted in Fig 1(b). It also reveals the relation between the BCP and the in-gap oscillations. The former arises from the gap closure due to the crossing of two unhybridized LLs, which is topologically protected while the latter is led by magnetic-field-induced modification of the energy gap when the e-h band hybridization is not strong enough to push the rest of LLs out of the gap.

***Correlation-driven spontaneous symmetry breaking.***—Further examining the temperature dependence in Fig. 2(c), we observe certain abnormalities at the transition point. By extracting the thermal activation gap at different $B_\perp$ as shown in Fig. 2(d), we found that the energy gap $\Delta$ reaches a dip around 12.5 T but does not entirely close. Several factors may contribute to the gap opening although its nature hasn't been well understood. Such gap closure at the topological phase transition is normally symmetry-protected. The spatial asymmetries including BIA and SIA are usually negligible in InAs/GaSb QWs [20]. But those effects really depend on the specific QW geometry and the external electric field, which may lead to a hybridization gap. On the other hand, electron-hole interaction may contribute to the gap opening as well since our sample live in the regime where single-particle hybridization and excitonic interaction are competing. And the spatial inhomogeneity caused by fluctuations in electrostatic potential or variations in the thickness of the QW can also attribute to the gap opening at the transition point. Similar phenomena that the gap doesn't close during the topological phase transition are also reported in similar InAs/InGaSb system [28] and two-dimensional material system [5].

A new mini-gap $\Delta^*$ reminiscent of the excitonic gap [8,9] is observed at low temperatures. It emerges at around 9 T and increases with the magnetic field. The different positions of the $\Delta$ dip and the $\Delta^*$ peak in the magnetic axis may indicate their distinctive mechanisms. It is likely that the $\Delta^*$ arises from excitonic binding between the electrons and holes around the crossing point. It may persist even in the weakly-noninverted regime as long as the binding energy is larger than the band gap [29–32], while the single-particle gap is supposed to close at the transition point (*i.e.*, ~13 T).

Next, we investigate the transition under different band inversions. Figure 3(b) and 3(e) show the bulk conductance maps as a function of $V_f$ and $B_\perp$ at $V_b = -1.5$ V and $V_b = 3$ V, respectively. At the CNP, different $V_b$ means different built-in electric fields and different degrees of the band inversion. At $V_b = -1.5$ V, where the bands are relatively shallowly-inverted, it shows an unambiguous phase transition from an inverted band insulator (characterized by the in-gap oscillations) to a new insulator with the square resistance up to 1 G$\Omega$. At $V_b = 3$ V, a deeply-inverted case, we observe a similar phase transition but with a widely broadening transition peak, which can be attributed to the weak hybridization and the band renormalization [15,16,33,34]. Comparing the magnetoconductance at the CNP, as shown in Fig. 3(a) and Fig. 3(d), the transition point for shallowly-inverted bands arises at a lower magnetic field than that in a deeply-inverted case, consistent with the phase transition picture.

Tuning the band inversion in InAs/GaSb system also changes the interplay of the interlayer Coulomb interaction and the tunneling. In the shallowly-inverted regime, the interaction effect is thought to be important to support a topological excitonic insulator [9]. We examine the in-plane magnetic field dependencies of the phase transition for different degrees of the band inversion. Figure 4(a) and 4(c) show the bulk magnetoconductance as a function of the tilt angle $\theta$ (defined in the inset) at $V_b = 2$ V and $V_b = -1.5$ V, respectively. In the deeply-inverted regime ($V_b = 2$ V), the BCP is found to increase with the in-plane magnetic field $B_{||} = B_{tot} \sin\theta$. For such a spatially separated e-h bilayer, applying $B_{||}$ is supposed to induce a relative momentum shift between two bands, $\Delta k = eB_{||}\langle z \rangle/\hbar$, where $\langle z \rangle$ is the vertical distance between two layers [35]. This may close the hybridization gap and convert the QSH insulator into a semimetal. According to the extracted peak conductance as a function of $B_{||}$ at $B_\perp = 14.4$ T in Fig. 4(b), the transition point can be considered a metal at $B_{||} = 13$ T where the BCP shows a rapid increase. The in-gap oscillations are also found to deviate from the original pattern under $B_{||}$. It indicates the in-plane field significantly changes the Fermi surface, which will not be affected if the gap is driven by excitonic interaction, further supporting the single-particle-hybridization-dominant picture in the deeply-inverted regime (more detailed explanation can be found in the SM).

An opposite trend of response to the in-plane magnetic field is found when the same device is tuned into the shallowly-inverted regime ($V_b = -1.5$ V), where the transition peak is continuously suppressed when the $B_{||}$ increases, and essentially disappears when the $B_{||} > 10$ T ($\theta > 71°$), as shown in Fig. 4(d). It indicates that the $B_{||}$ converts the gap-closing topological phase transition into a smooth crossover without gap closing. Such a crossover between two topologically distinctive phases has to be accompanied by symmetry breaking [5,10]. Here the $B_{||}$ component reduces the interlayer tunneling and hence enhances the effective interaction strength U/A. Overall, the conductance suppression under $B_{||}$ in shallowly-inverted InAs/GaSb bilayer suggests a correlation-driven state with spontaneous symmetry breaking before the transition. The robustness of the in-gap oscillations in a lower field supports this assertion as well.

*Discussion and conclusion.*—The remainder of this Letter will focus on more discussions on the exciton effect. For a dilute e-h system, long-range Coulomb interaction prefers to pair electrons and holes into a bosonic exciton condensate. The excitonic gap is weakly dependent on the momentum shift induced by an in-plane magnetic field and to some extent it is enhanced due to a suppression of e-h interlayer tunneling. We have observed two well-separated regimes which can be parametrized by the degree of the band inversion. For the deeply-inverted regime, as shown in Fig. 4(a) and 4(b), the gap closing at the transition point can be well explained within a single-particle model. For the shallowly-inverted regime in Fig. 4(c) and 4(d), a smooth crossover without gap closing suggests a correlation-driven state with spontaneous symmetry breaking before the transition. All of these observations are consistent with the excitonic insulator instability mechanism in InAs/GaSb bilayer when considering the interplay between the Coulomb interaction and the hybridization [9,36]. More recently, a time-reversal-symmetry-broken state is theoretically proposed for InAs/GaSb bilayers [12], which emerges between the non-inverted bands and the inverted bands where QSH insulators are observed. This regime exactly falls into the dilute limit where excitonic insulators [9,37]and the results in Fig. 4(c) and 4(d) are observed.

In conclusion, for the first time, we observed a magneto-driven topological phase transition from a QSHI to a NI. Significantly, as the e-h system is turned towards shallowly-inverted regime, long-range Coulomb interaction dominates and converts the topological phase transition into a smooth crossover without gap closing. It suggests a correlated state with spontaneous symmetry breaking. Our work demonstrates a highly controllable e-h system under high magnetic fields to explore the interplay between topology and interaction.

*Acknowledgments.*—We gratefully acknowledge C. L. Yang and P. L. Li for technical assistance. The work at PKU was funded by the National Key R&D Program of China (Grants No. 2017YFA0303300 and 2019YFA0308400), and by the Strategic Priority Research Program of Chinese Academy of Sciences (Grant No. XDB28000000). The InAs/GaSb quantum wells structures were prepared by molecular beam epitaxy by Gerard Sullivan. A portion of this work was performed at the National High Magnetic Field Laboratory, which was supported by the National Science Foundation Cooperative Agreement No. DMR-1157490 and the State of Florida. The work at UCAS was supported by the National Key R&D Program of China (Grant No. 2018YFA0305800) and the National Natural Science Foundation of China (Grant No. 12174387).

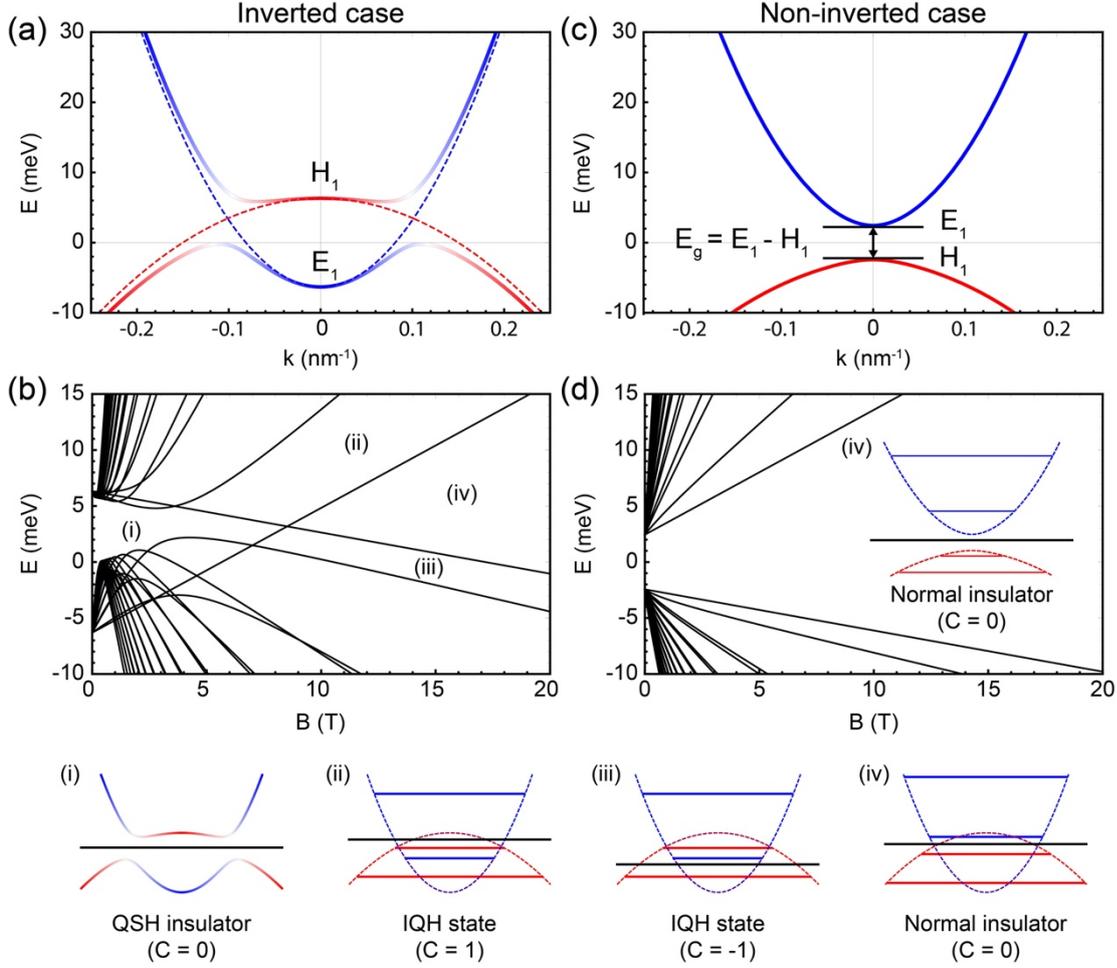

**Fig. 1.** (a) Energy dispersion of type-II InAs/GaSb QWs with an inverted band structure based on a non-interacting model. The dashed lines describe the dispersion without hybridization, where E1 and H1 denote the electron and hole bands, respectively. (b) LL spectra for the inverted case in (a). Two unhybridized LLs divide the region in the vicinity of the inversion transition into four regimes with different Chern numbers labeled (i) to (iv). Corresponding schematics of the band structure are plotted below: (i) a QSHI state with an inverted band order with the Chern number $C = 0$; (ii) a QH state with only the lowest electron LL occupied ($C = 1$); (iii) a QH state with only the lowest hole LL occupied ($C = -1$); (iv) a QH states with no occupation in LLs ($C = 0$), namely, a normal insulator, where the blue (red) lines denote the electron (hole) LLs and the black lines represent the chemical potentials. (c) Energy dispersion and (d) LL spectra for a non-inverted case.

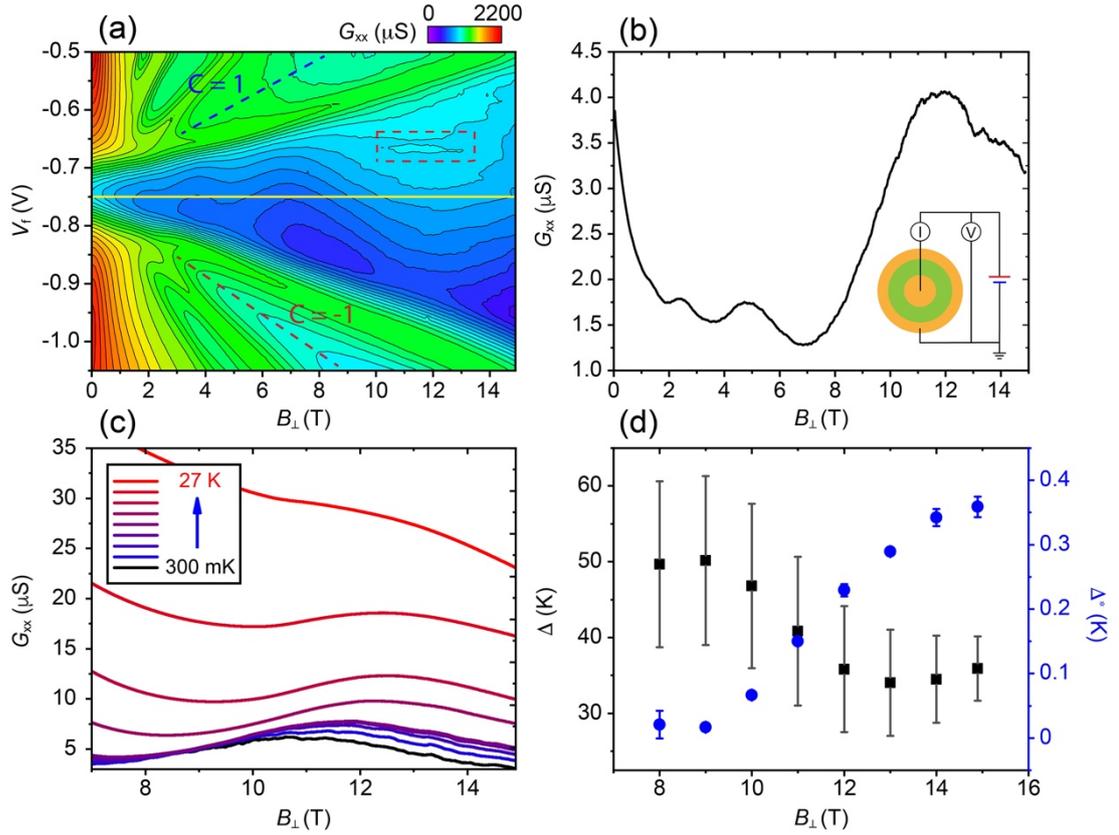

**Fig. 2.** (a) Bulk conductivity as a function of $V_f$ and $B_\perp$ at $V_b = 0$ V. The conductivity is measured from a dual-gated Corbino device. The red dashed-line box marks a bulk conductance peak around 12 T. The blue (red) dashed line denotes the IQH state with Chern number $C = 1$ ($C = -1$) as described in Fig. 1(b). The magneto-conductance trace is plotted in (b) where the Fermi energy is pinned at the CNP ($V_f = -0.75$ V, yellow line in the contour plot). (c) The temperature dependence of the bulk conductance peak near the transition point at the CNP. (d) The energy gap vs perpendicular magnetic field acquired by the Arrhenius analysis from the data in (c). The energy gap $\Delta$ shows a minimum around 13 T accompanied by the opening of a small gap $\Delta^*$.

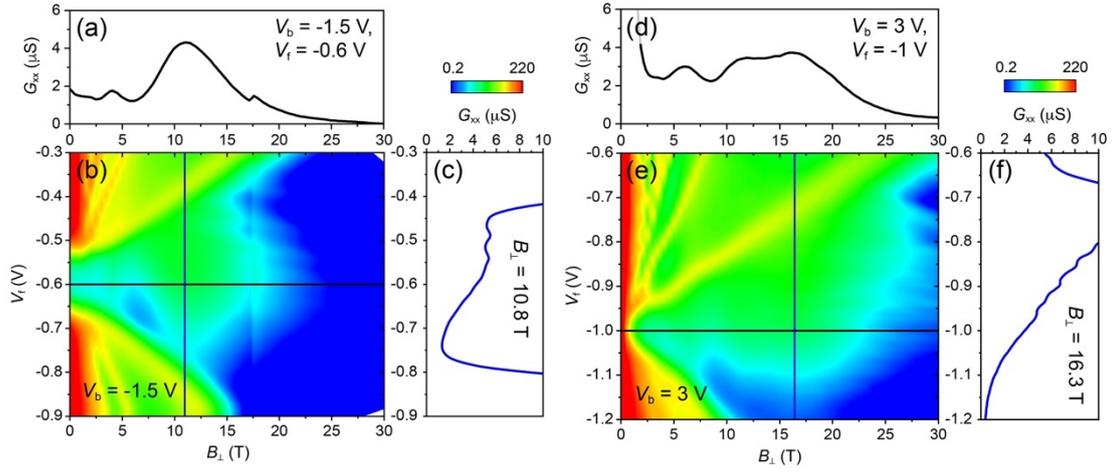

**Fig. 3.** The magneto-induced phase transitions under different band inversions. Two respective bulk conductivity maps are plotted at (b) $V_b$ = -1.5 V and (e) $V_b$ = 3 V. Corresponding conductance traces (horizontal linecuts) extracted with the Fermi energy pinned at the CNP are shown (a) at $V_f$ = -0.6 V (n = p ~ 1.24 × $10^{11}$ $cm^{-2}$) and (d) at $V_f$ = -1 V (n = p ~ 2.48 × $10^{11}$ $cm^{-2}$), respectively. Vertical linecuts are plotted to present bulk conductance at the transition point (c) at $B_\perp$ = 10.8 T and (f) at $B_\perp$ = 16.3 T.

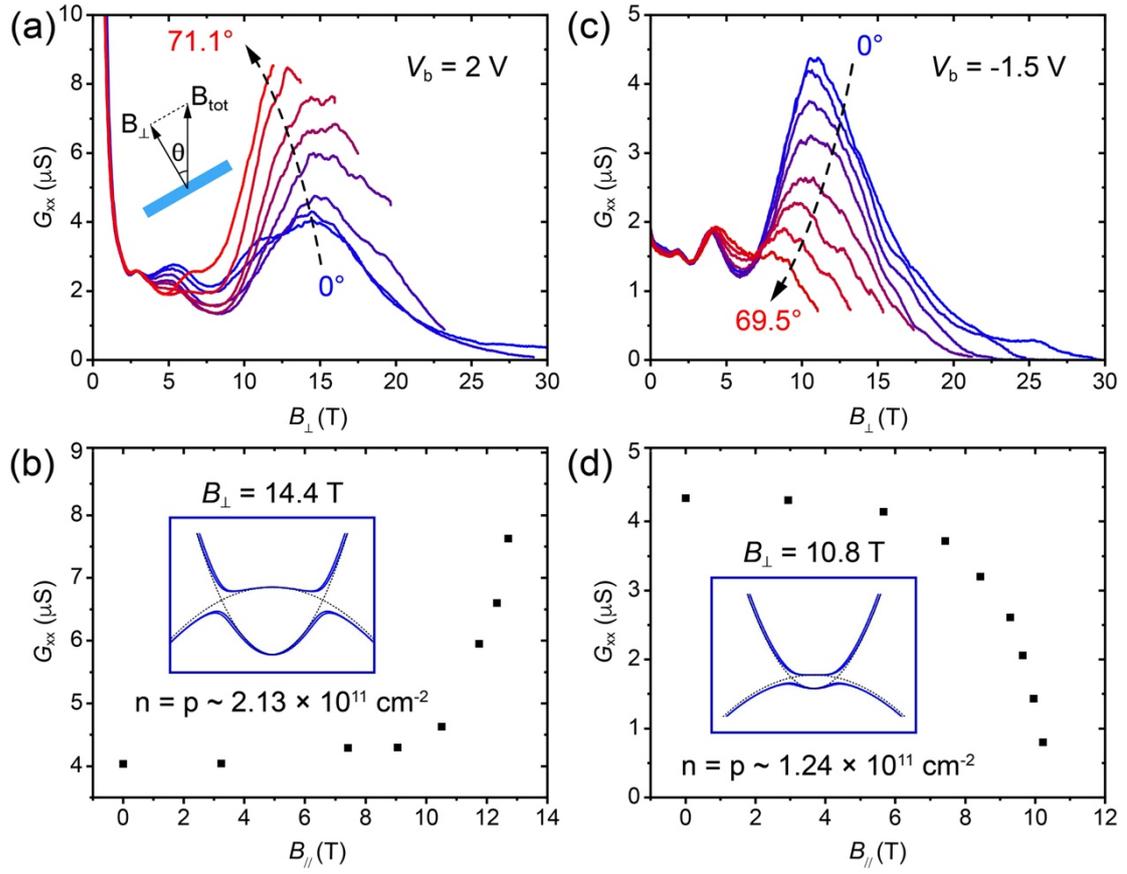

**Fig. 4.** The in-plane magnetic field dependence (a) at $V_b = 2$ V (deeply-inverted regime), and (c) at $V_b = -1.5$ V (shallowly-inverted regime). The corresponding peak conductance is extracted with $B_\perp$ fixed (b) at 10.8 T ($V_b = -1.5$ V) and (d) at 14.4 T ($V_b = 2$ V).

# Supplemental Materials for Magneto-Induced Topological Phase Transition in Inverted InAs/GaSb Bilayers

*Zhongdong Han,*[1,*,#] *Tingxin Li,*[1,**] *Long Zhang,*[2] *and Rui-Rui Du*[1,3,#]

[1]International Center for Quantum Materials, School of Physics, Peking University, Beijing 100871, China

[2]Kavli Institute for Theoretical Sciences and CAS Center for Excellence, University of Chinese Academy of Sciences, Beijing 100190, China

[3]Collaborative Innovation Center of Quantum Matter, Beijing 100871, China

[*] *Present address: Laboratory of Atomic and Solid State Physics, Cornell University, NY 14850, USA.*

[**] *Present address: Key Laboratory of Artificial Structures and Quantum Control (Ministry of Education), School of Physics and Astronomy, Shanghai Jiao Tong University, Shanghai, 200240, China.*

[#] Email: zh352@cornell.edu

Email: rrd@pku.edu.cn

**Wafer structure**

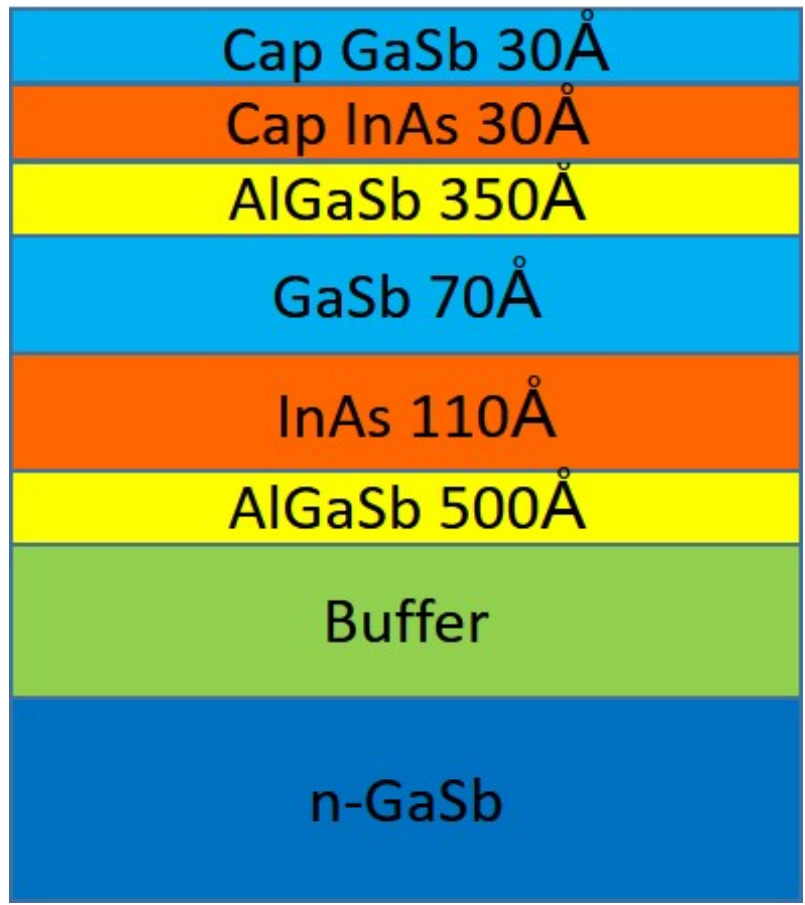

**Fig. S1.** Wafer structure. Our high-quality QW samples were grown by semiconductor molecular beam epitaxy technique on an N-GaSb (001) substrate, which also severs as a back gate. A 7 nm GaSb layer and an 11 nm InAs layer are sandwiched by two AlGaSb barriers after growing a 1 um AlGaSb buffer layer. Cap layers on the top are designed for protection from oxidation and for pining the initial carrier density.

## Zero-field conductance and carrier density extracted from SdH oscillations

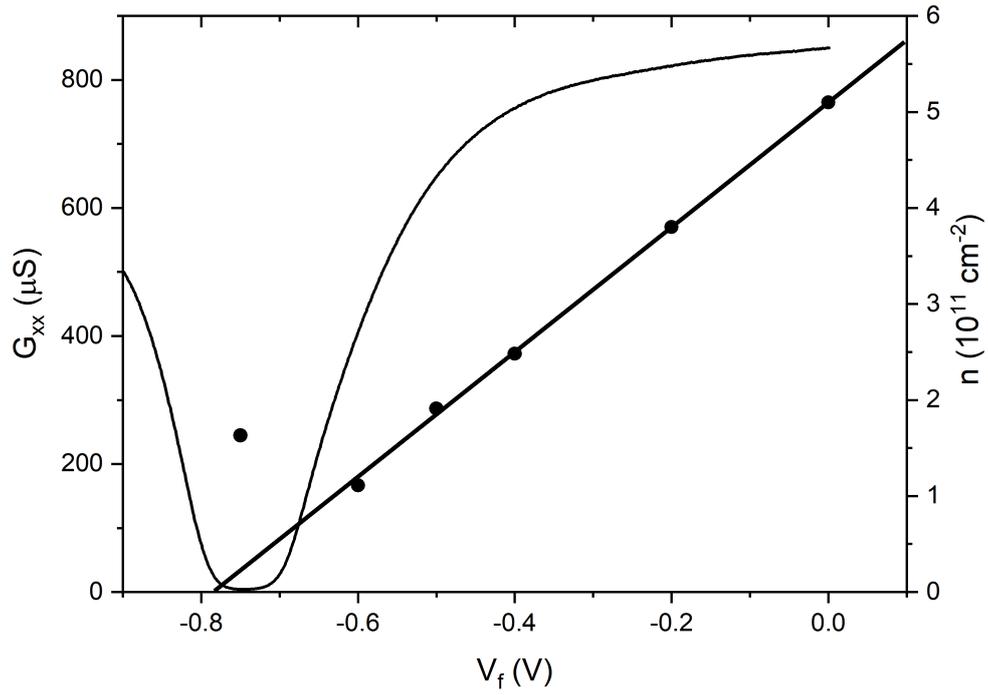

**Fig. S2.** Bulk conductance $G_{xx}$ as a function of the front gate voltage $V_f$ at $V_b = 0$ and $B = 0$ T. The density n is extracted from SdH oscillations and in-gap oscillations. It shows a linear dependence outside the gap and a sudden jump into the gap, which can be explained by the band renormalization by including the quasiparticle self-energy. The black line is a guide by the eye.

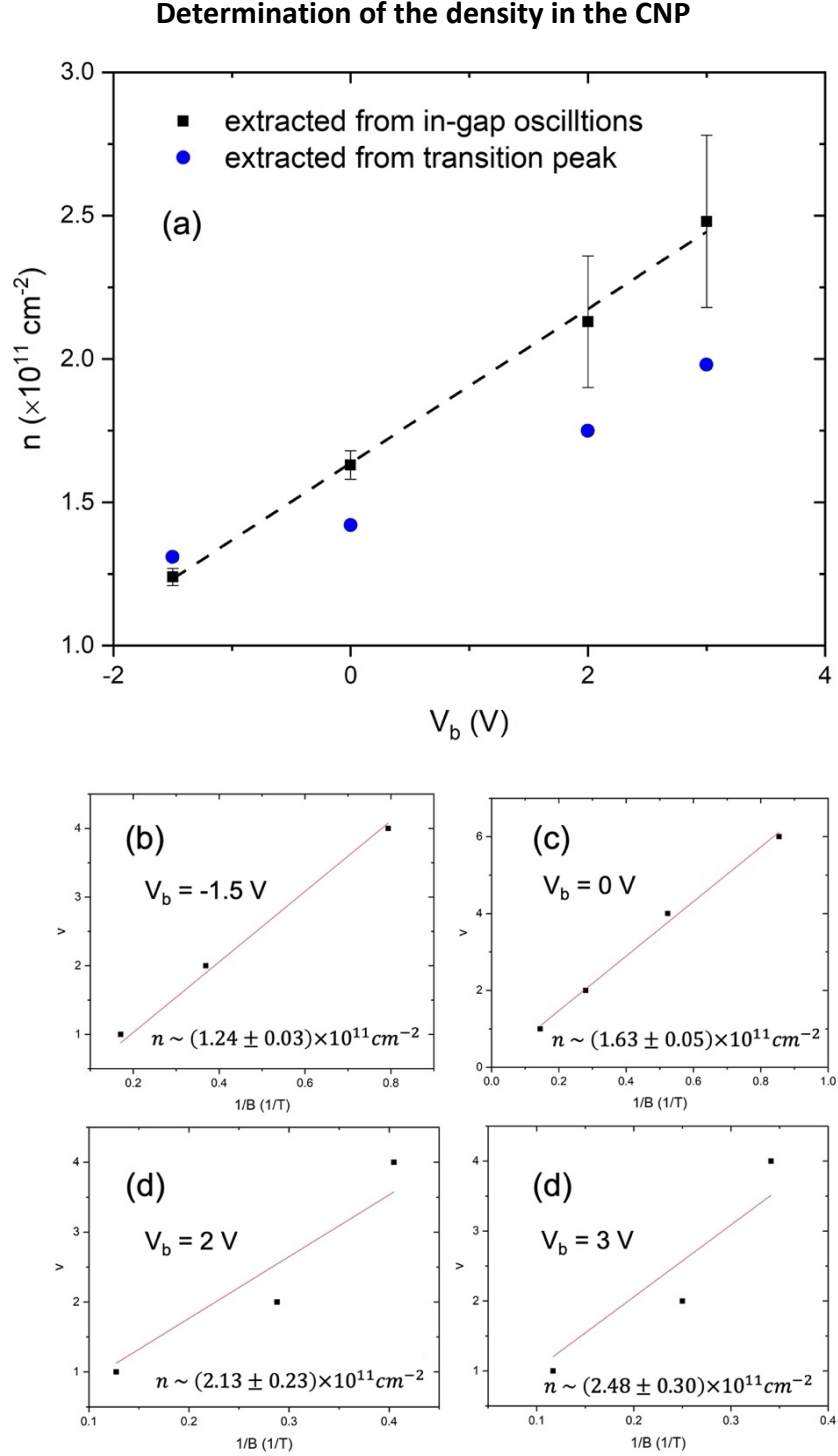

**Fig. S3.** (a) Electron/hole density at the CNP, which characterizes the degree of the band inversion. Black squares are extracted from the in-gap oscillations. Blue dots are calculated from the transition peak where both electrons and holes occupy half of the lowest LLs. The black dashed line indicates that the gate dependence of $n_{cross}$ is quite linear. Corresponding filling factor index diagrams are plotted (b) at $V_b$= -1.5 V, (c) at $V_b$= 0 V, (d) at $V_b$= 2 V, and (e) at $V_b$= 3 V.

**Arrhenius plots for gap fitting at high magnetic field**

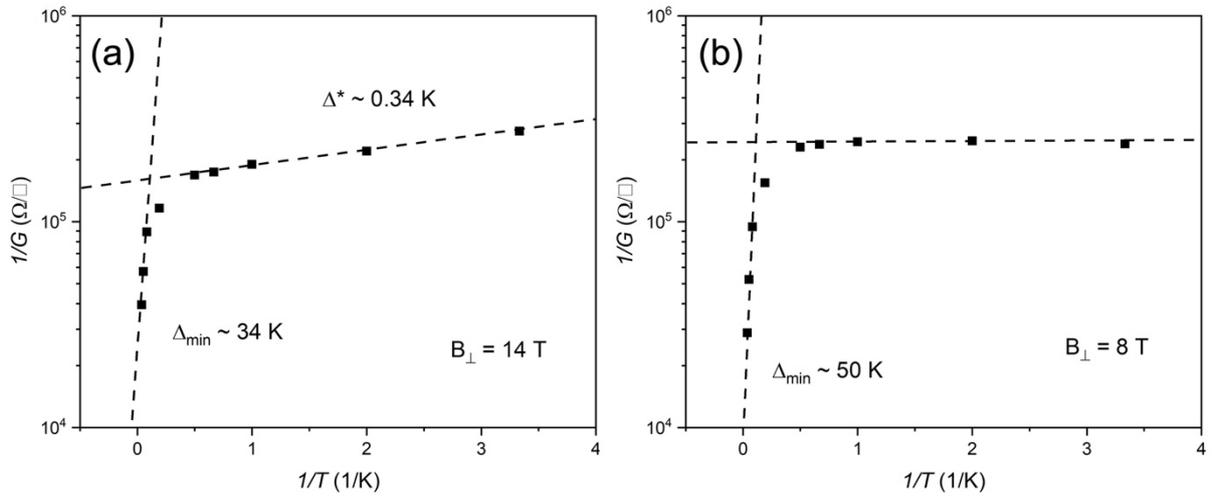

**Fig. S4.** Two typical Arrhenius plots are shown (a) at $B_\perp = 14$ T and (b) at $B_\perp = 8$ T. The fitting gaps $\Delta$ and $\Delta^*$ are plotted in Fig. 2(d) in the main text.

**In-gap magneto-conductance after subtracting the background**

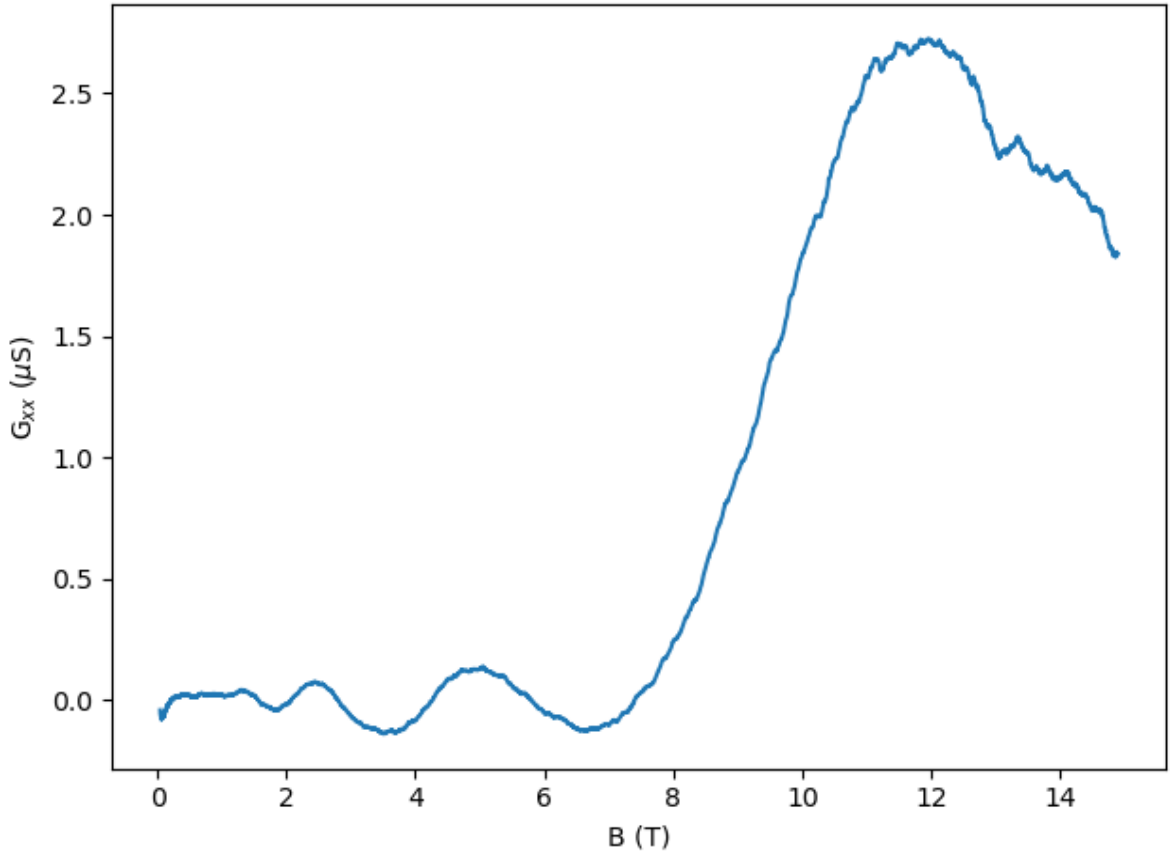

**Fig. S5.** The magneto-conductance trace at $V_f = -0.75$ V and $V_b = 0$ V after removing the background. The raw data is plotted in Fig. 2(b) in the main text.

**Device geometry**

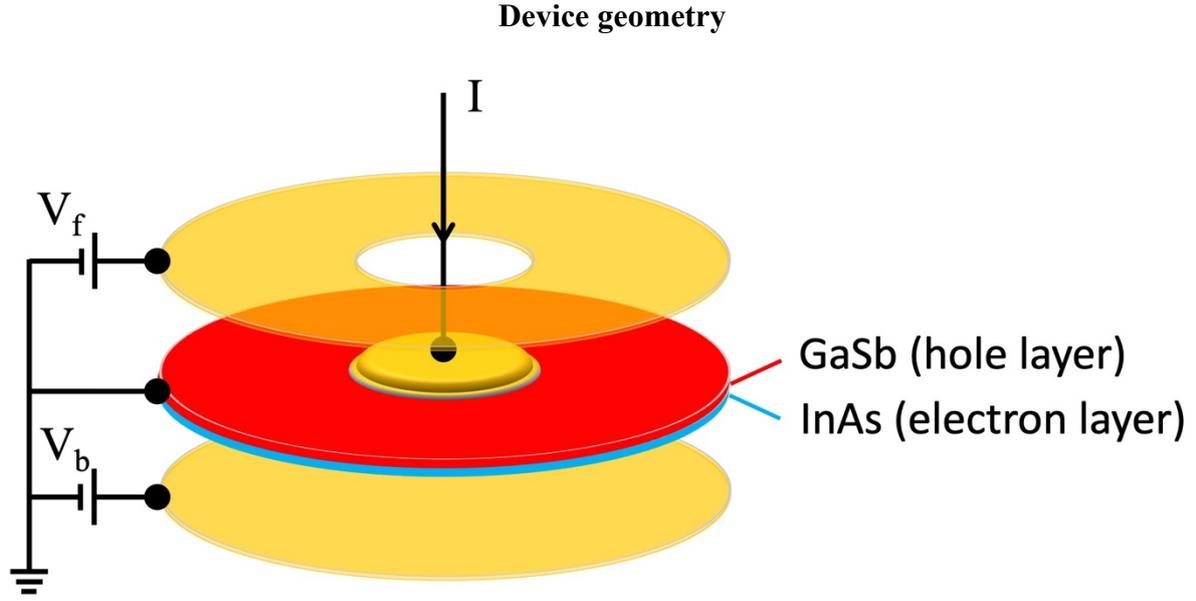

**Fig. S6.** Device geometry. In our experiment, we use a dual-gated Corbino device to study the magneto-transport properties of the bulk. The n-doped substrate serves as a back gate. And 10 nm Ti/ 100 nm Au are deposited as a front gate after growing an $Al_2O_3$ dielectric layer. In this geometry, the edge contribution can be excluded because the edge states are shunted by the metal electrodes. And a dual-gated configuration allows us to independently tune the density and the out-of-plane electric field.

**The position of the Fermi energy relative to the LLs in a high magnetic field**

A common practice to describe the position of the Fermi energy relative to the LLs is using the filling factor $\nu = \frac{nh}{eB}$, where n is the carrier density for a two-dimensional electron gas and B is the perpendicular magnetic field [1]. It describes how many LLs are occupied under the magnetic field. As the degree of degeneracy of one LL is $D = \frac{h}{eB}$, the filling factor can be easily derived by $\nu = \frac{n}{D} = \frac{nh}{eB}$ (see Fig. S7(a)). If $\nu$ is an integer, which means $\nu$ LLs are fully occupied and the $(\nu + 1)$-th LL is empty, the Fermi energy is inside the gap between the $\nu$-th and the $(\nu + 1)$-th LLs. If $\nu$ is between two integers $m < \nu < m + 1$, which means m LLs are fully occupied and the $(m + 1)$-th LL is partially filled, the Fermi energy is in principle pinned at the $(m + 1)$-th LL. In fact, the Fermi energy may deviate from the $(m + 1)$-th LL by considering extended and localized states. For a single carrier system with fixed density n in two dimensions, as the magnetic field increases, the degree of degeneracy $D$ of each LL is increasing and the filling factor $\nu$ is decreasing. When the Fermi energy is in the gap between two LLs, the bulk is insulating and the Gxx is small. But every time the Fermi energy passes one LL, the bulk becomes conducting and the Gxx shows a peak. It's the origin of the Shubnikov de-Hass oscillation in a two-dimensional metal.

Before we move to the two-carrier case, we need to figure out the real position of the Fermi energy for the single-carrier case. For simplicity, we take an ideal two-dimensional electron gas (2DEG) system as an example and spin degeneracy is not considered here. As we already know the relative position of the Fermi energy to the LLs, the Fermi energy is always pinned at the $(m + 1)$-th LL when the filling factor is between two integers $m < \nu < m + 1$. As the magnetic field increases, the Fermi energy will suddenly drop to the $m$-th LL when $\nu < m$. The energy of the $m$-th LL is $E_m = (m - \frac{1}{2})\hbar \frac{eB}{m_e^*}$, where $m_e^*$ is the effective mass for the electron. Under a specific magnetic field where the filling factor $\nu$ is slightly less than $m$, the Fermi energy is pinned at the $m$-th LL, $E_f = \left(m - \frac{1}{2}\right) \hbar \frac{eB}{m_e^*} = \left(\frac{nh}{eB} - \frac{1}{2}\right) \hbar \frac{eB}{m_e^*} = \frac{2\pi n \hbar^2}{m_e^*} - \frac{e\hbar}{2m_e^*} B = E_{f0} - \frac{e\hbar}{2m_e^*} B$, where the first term is the Fermi energy at zero field $E_{f0} = \frac{2\pi n \hbar^2}{m_e^*}$, and the second term is half the cyclotron energy $\frac{1}{2}\hbar\omega_c = \frac{e\hbar}{2m_e^*} B$. As the magnetic field increases, the Fermi energy will increase linearly with the magnetic field until the filling factor $\nu < m - 1$. Right before the Fermi energy drops to the $(m - 1)$-th LL—in other words, $\nu$ is slightly greater than $m - 1$ and the Fermi energy is still pinned at the $m$-th LL—the highest energy that can be reached is $E_f = \left(m - 1 + \frac{1}{2}\right) \hbar \frac{eB}{m_e^*} = \left(\frac{nh}{eB} + \frac{1}{2}\right) \hbar \frac{eB}{m_e^*} = E_{f0} + \frac{e\hbar}{2m_e^*} B$. So, the Fermi energy will oscillate around $E_{f0}$ with increasing amplitude $\frac{e\hbar}{2m_e^*} B$, as shown in Fig. S7(b). At certain magnetic field $B_\nu = \frac{nh}{e\nu}$ where $\nu$ is an integer, the Fermi energy suddenly drops from $E_{f0} + \frac{e\hbar}{2m_e^*} B_\nu$ to $E_{f0} - \frac{e\hbar}{2m_e^*} B_\nu$.

Then let's consider an independent electron-hole bilayer system where electrons and holes have the same density $n_0$. Notice that the degree of degeneracy of the LL is independent of the type of the carrier. The electrons and the holes share the same filling factor $\nu = \frac{n_0 h}{eB}$ as they have the same density. As they are independent, we can plot their respective Fermi energy, as shown in Fig. S7(c). The blue and red lines respectively represent the Fermi energies for electrons and holes. They have the same oscillation frequency but different amplitudes due to different effective masses and $\pi$ phase difference due to the opposite signs of the mass.

While in our realistic system, electrons and holes are hybridized by interlayer tunneling and are shorted through contacts (see Fig. S6). Because they are electrically shorted, electrons and holes share the same Fermi energy. In Fig. S7(c), we find there are four different regions.

1) When the filling factor is an integer, the Fermi energies for both electrons and holes are in the middle of the gap between two LLs. It's equal to the Fermi energy at zero magnetic field $E_{f0}$.

2) When the filling factor is greater than a half integer (the region where the red line is higher than blue line in Fig. S7(c)), the Fermi energy of holes is higher than the Fermi energy of electrons. But electron and hole layers are shorted. They share the same Fermi energy. As shown in Fig. S7(d), electrons in the highest occupied LL in the hole layer (higher energy) will move to electron layer (lower energy) to lower the energy and fully occupied the highest occupied LL in electron layer. Then the Fermi energies in two layers are equal and locate at the gap between these two LLs.

3) Similar situation occurs when filling factor is less than a half integer. As shown is Fig. S7(e), the Fermi energy of electrons is higher than the Fermi energy of holes. Then electrons will move from electron layer to hole layer and make the system insulating again. The Fermi energy for both layers is in inside the gap between these two LLs.

4) A special case is when the filling factor is at a half integer (where red and blue lines cross at $E_{f0}$). In this case, the highest occupied LLs for both electron and hole layers locate at the same energy $E_{f0}$ and they are both half filled. While the interlayer tunneling pushes these two LLs away from $E_{f0}$ and open a hybridization gap as shown in Fig. S7(f). The system becomes insulating again with the Fermi energy locating inside the hybridization gap. We would like to note that there is an exception when the filling factor is equal to ½. This crossing point is topologically protected by certain symmetries such as the spatial mirror symmetry [2,3]. It's the critical point of the topological phase transition from a quantum spin Hall insulator (QSHI) to a normal insulator (NI) with a gap closure.

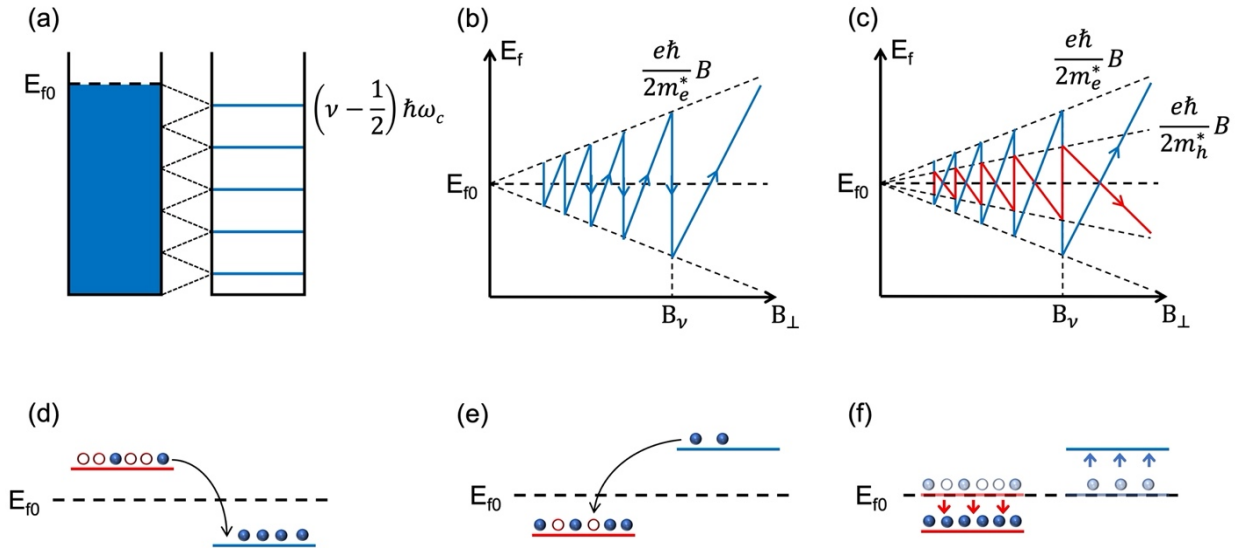

**Fig. S7.** (a) Landau quantization of a 2DEG system under a uniform magnetic field. $E_{f0}$ denotes the Fermi energy at zero magnetic field. These blue lines represent the discrete LLs with quantized cyclotron energy $E_\nu = (\nu - \frac{1}{2})\hbar\omega_c$, where $\omega_c = \frac{eB}{m_e^*}$ is the cyclotron

frequency. The oscillations of the Fermi energy (b) for a single-carrier system with fixed density and (c) for an independent electron-hole bilayer system where the densities of electrons and holes are equal. The oscillation amplitudes linearly increase with the magnetic field depending on the effective mass. Schematic diagrams of charge transferring and the chemical potential balancing for a coupled electron-hole bilayer system (d) when the highest occupied LL of hole layer (red line) is higher than that of electron layer (blue line), (e) when the highest occupied LL of hole layer is lower than that of electron layer, and (f) when the highest occupied LL of hole layer locates at the same energy with that of electron layer.

## The phase diagram for InAs/GaSb system

In single-particle picture (see Fig. S8(a)), the phase diagram for InAs/GaSb system is divided into six regimes: 1) two-dimensional topological insulator (2DTI) when the electron and hole bands are inverted and the Fermi energy locates within the gap; 2) NI when two bands are not inverted and the Fermi energy locates within the gap; 3) n-doped 2DTI; 4) p-doped 2DTI; 5) n-doped NI, and 6) p-doped NI. The dashed line indicates the boundary between an inverted band and a noninverted band. And the orange dot denotes the critical point of the topological phase transition between a 2DTI and a NI. More details can be found in Ref. [4].

In Fig. S8(a), the gap must close at the critical point during the topological phase transition. The gap closure is topologically protected under the same symmetry. Recent work [5] reported that this gap doesn't close during the gate-controlled topological phase transition, which indicates the correlation-driven spontaneous symmetry breaking [6-9]. The combination of transport measurements and terahertz transmission spectroscopy measurements [10] further point out that excitonic interaction exists in the dilute limit in InAs/GaSb bilayer system. Thus, under a many-body interaction picture (see Fig. S8(b)), there exist an excitonic insulator (EI) state near the critical point if the binding energy exceeds the band gap. In our experiment, the excitonic interaction is even enhance by applying a perpendicular magnetic field, which makes the electron and hole bands flatter and more symmetric.

If we apply a large enough in-plane magnetic field, the interlayer tunneling is suppressed and single-particle hybridization gap will close. It turns a 2DTI into a semimetal (SM). However, an interaction-driven EI will not be influenced by in-plane magnetic fields. The blue pentagram in Fig. S8 indicates the specific region where we are in this device. It's a regime where the interlayer tunneling and Coulomb interaction effects coexist. If we are more on the single-particle tunneling side, the gap tends to close (see Fig. 4a in the main text). If we are more on the many-body interaction side, the gap is enhanced because the tunneling is prohibited (see Fig. 4c in the main text.)

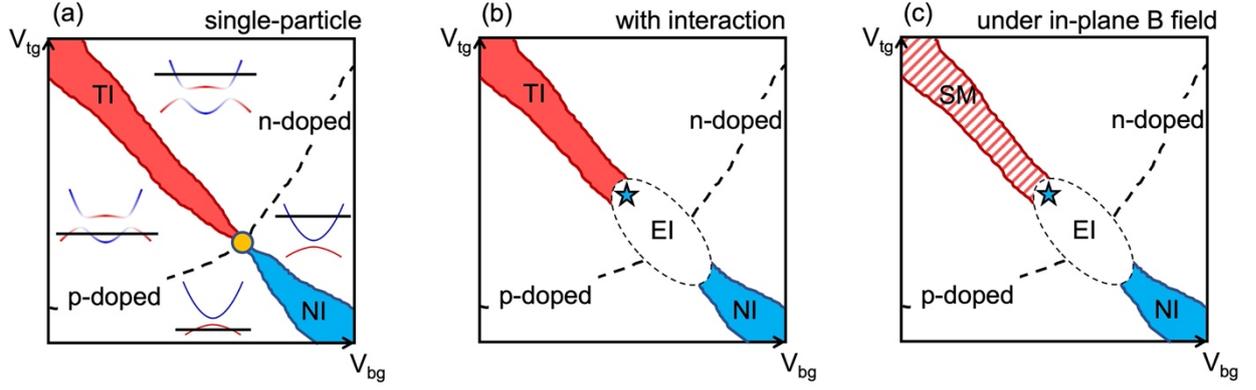

**Fig. S8.** The phase diagram for InAs/GaSb bilayer system (a) under a single-particle picture, (b) under a many-body interaction picture, and (c) under an in-plane magnetic field. The dashed line in (a) denotes the boundary between an inverted band and a noninverted band. On the inverted side, when the chemical potential is tuned into the gap, it's a 2DTI (red region). On the other side, it's a NI (blue region). The orange dot denotes the critical point of the topological phase transition between these two phases. Under an interaction picture in (b), an EI state emerges near the critical point. The blue pentagram indicates the regime where we are around in our experiment. After applying an in-plane magnetic field, the interlayer tunneling is suppressed. It turns a 2DTI state into a SM state, as denoted by the red dashed region in (c).

**Effect of in-plane magnetic field on the in-gap oscillation**

Within the hybridization picture, the in-plane magnetic field effect on the in-gap oscillation can be considered from two limits. In the zero-field limit, the tunneling between electron and hole layers results in a hybridization gap at the crossing points of two bands at finite k. Although there is no Fermi surface inside the gap, the unique inverted-band structure still gives rise to quantum oscillations with a frequency determined by the enclosed area of the band edge. This part has been thoroughly discussed in our previous work [11] and has been widely reported in similar systems [12-14].

While in the high-in-plane-magnetic-field limit, the electron and hole bands are shifted in k space by $\Delta k = eB_{||}\langle z \rangle / \hbar$, where $\langle z \rangle$ is the spatial distance between two layers. In this limit, the in-plane field is strong enough to fully decouple the electron and hole bands in the momentum space, resulting a semi-metallic system. The semimetal still exhibits quantum oscillations, with the frequency determined by the area of the Fermi surface for both electron and hole pockets, which is close to the oscillation frequency in the zero-field case.

When the in-plane magnetic field is in between these two limits, the electron and hole bands are shifted in k space but not fully decoupled. It becomes a semi-metal, but the Fermi surface is smaller than the fully decoupled case. In this case, the frequency of the quantum oscillations, which is determined by the area of the Fermi surface, also gets reduced.

In our experiments, we fixed the tilt angle θ of the magnetic field instead of the in-plane component. When we increase the out-of-plane magnetic field along x-axis in Fig. 4(a) in the main text, the in-plane magnetic field $B_{||} = B_{\perp} \tan\theta$, also increase from zero. At large tilt angles, the system will go through from the zero-field limit to high-field limit. It might mess up the oscillation phases due to the change of frequency during the transition between two limits. It refers to a peak-to-dip transition (π phase shift) in Fig. 4(a) in our experiments. This deviation of the quantum oscillation at large tilt angle is consistent with the single-particle picture where the hybridization effect is dominant.

## Determination of Chern numbers by Streda formula

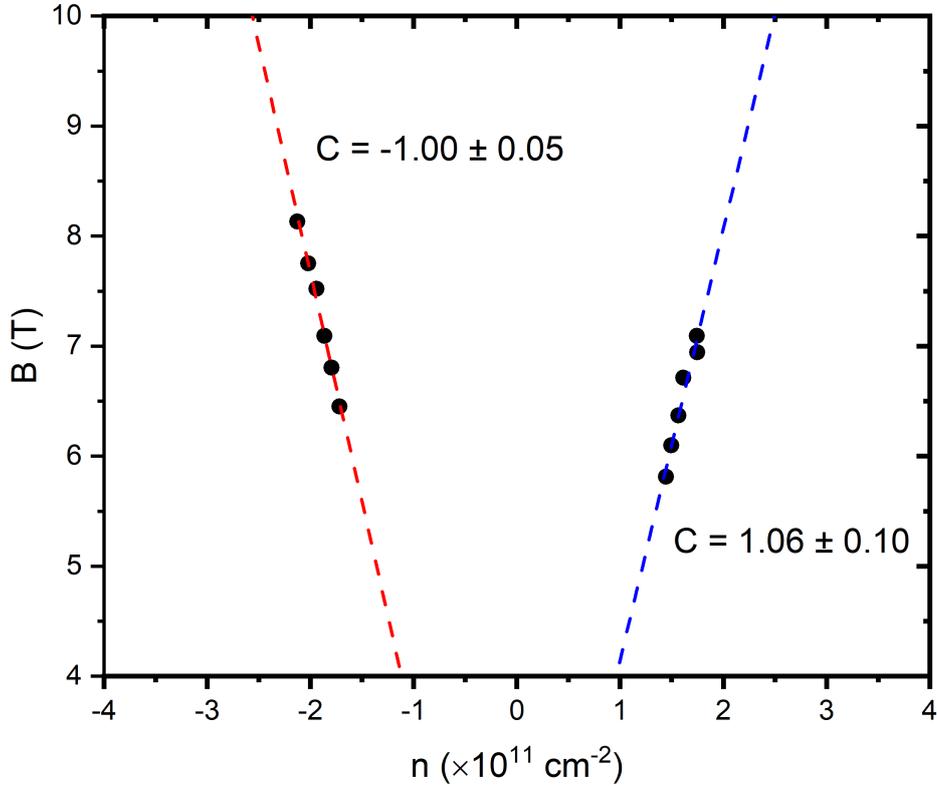

**Fig. S9.** Determination of Chern numbers by streda formula C = $\frac{h}{e}\frac{\Delta n}{\Delta B}$, where the density n is determined by SdH oscillation. The dispersions of these two IQH states in Fig. 2(a) in the main text in doping density and magnetic field verify that their Chern number are $1.06 \pm 0.1$ and $-1.00 \pm 0.05$, consistent with the phase diagram shown in Fig. 1(a) in the main text.

## Exciton effect under a high perpendicular magnetic field

We would like to briefly comment on InAs/GaSb bilayer under a perpendicular magnetic field. At the transition point, the physics can be described by interactions between two partially filled electron and hole LLLs, which can be viewed as an ideal flat band system. Due to confinement effects and reduced screening in low dimensions, the highly degenerate e-h gases under a high magnetic field favor an excitonic condensate state [15,16]. It's then interesting to compare the electron-hole bilayer in InAs/GaSb with the electron-electron (e-e) bilayer in GaAs/InGaAs [17] or graphene [18,19]. While in the latter case with each electron layer occupying a half-filled LLL, the QH excitonic condensation is the ground state and the ratio between the separation of two layers d and the magnetic length $l_B = \sqrt{\hbar/eB}$ is an important parameter to characterize the exciton binding strength. When $d/l_B \lesssim 1.8$, excitons start to condense with a long-range macroscopic order for the bilayer electron gas system [20,21]. For our e-h system, $d/l_B = 1.14$ is in the shallowly-inverted regime ($V_b$ = -1.5 V) and reaches 1.32 in the deeply-inverted regime ($V_b$ = 2 V). The boundary to support an excitonic gap is $1.14 \leq d/l_B \leq 1.32$, smaller than that in the e-e case. Moreover, we note that the following aspects may become important when considering the physics of e-h bilayers: 1) in the charge sector, the interlayer Coulomb attraction which drives excitonic pairing may dominate over intralayer interactions, and therefore the excitonic condensation may possess different characteristics compared to the QH excitonic condensation; and 2) the effective electron mass in InAs layer increases to $m_e^* = (0.032 + 0.005 \times B)m_e = 0.086m_e$ [22] at the transition point (B = 10.8 T), getting close to the effective hole mass ($m_h^* = 0.136m_e$). With heavier mass and more symmetric bands, the e-h bilayer system is expected to realize the exciton condensation at a considerably higher temperature [23]. 3) In the spin sector, occupied electron and hole states in these two spin-resolved LLLs carry spins in the same direction, which is opposite in the e-e case considering the particle-hole transformation. Compared with the base temperature $k_BT \sim 0.03$ meV, the Zeeman splitting is $g_e\mu_B B_c = 7.18 meV$ for electrons and $g_h\mu_B B_c < 1.88\ meV$ for holes [22]. Besides, the hybridization between LLs further increases the singlet-triplet exchange energy to $\Delta = $ 9~11 *meV* [22,24]. The triplet interaction with this spin texture favors the Bose-Einstein condensation of p-wave excitons [25-29].